\documentclass[%
reprint, superscriptaddress,
 amsmath,amssymb,
 aps, prl,
floatfix,
]{revtex4-2}
\usepackage{graphicx}
\usepackage{amsmath}
\usepackage{hyperref}
\def\refjnl#1{{\rmfamily#1}}%
\renewcommand\apj{\refjnl{ApJ}}%    % Astrophysical Journal 
\newcommand\apjl{\refjnl{ApJL}}     % Astrophysical Journal, Letters 
\newcommand\apjs{\refjnl{ApJS}}%    % Astrophysical Journal, Supplement 
\newcommand\pasp{\refjnl{PASP}}%     % Publications of the ASP 
\newcommand\mnras{\refjnl{MNRAS}}%   % Monthly Notices of the RAS 
\newcommand\aap{\refjnl{A\&A}}%     % Astronomy and Astrophysics 

\def\software#1{\vskip 6pt{
\frenchspacing
\font\foo=cmr10
\fontdimen2\foo=3pt %% Changed from 1.5pt to 3pt, March12, 2019
{\large \it Software: }
#1
\fontdimen2\foo=3.33333pt
}
}

\begin{document}

\title{Constraining Nuclear Symmetry Energy with Multi-messenger Resonant Shattering Flares}

\author{Duncan Neill}
\affiliation{Department of Physics\\
University of Bath \\
Claverton Down \\
Bath, United Kingdom}

\author{Rebecca Preston}
\affiliation{Dept. of Physics \& Astronomy \\
Texas A\&M University at Commerce  \\
Commerce, Texas, USA}

\author{William G. Newton}
\affiliation{Dept. of Physics \& Astronomy \\
Texas A\&M University at Commerce  \\
Commerce, Texas, USA}

\author{David Tsang}
\affiliation{Department of Physics\\
University of Bath \\
Claverton Down \\
Bath, United Kingdom}

%% Mark off the abstract in the ``abstract'' environment. 
\begin{abstract}

Much effort is devoted to measuring the nuclear symmetry energy through neutron star (NS) and nuclear observables. Since matter in the NS core may be non-hadronic, observables like radii and tidal deformability may not provide reliable constraints on properties of nucleonic matter.
We demonstrate that coincident timing of a resonant shattering flare (RSF) and gravitational wave signal during binary NS inspiral probes the crust-core transition region and provides constraints on the symmetry energy comparable to terrestrial nuclear experiments. We show that nuclear masses, RSFs and measurements of NS radii and tidal deformabilities constrain different density ranges of the EOS, providing complementary probes.

\end{abstract}


\maketitle

%% MAIN PAPER STARTS HERE:

\section{Introduction} \label{sec:intro}

The distinction between protons and neutrons (isospin) is one of the most significant degrees of freedom affecting nuclear interactions on the hadronic level. In bulk nuclear matter, the effect of exchanging neutrons and protons is encapsulated by the difference in the binding energy of symmetric nuclear matter (SNM) and pure neutron matter (PNM): the nuclear symmetry energy.
This energy is typically expanded around nuclear saturation density (where most symmetric nuclei lie, $n_{\rm s}\approx0.16 \text{ fm}^{-3}$) to obtain a set of (isovector) parameters, describing its magnitude ($J$), slope ($L$), curvature ($K_{\rm sym}$), etc. %and so on.
Many experiments have probed the effect of isospin asymmetry (e.g. giant dipole resonances \citep{trippa2008giant}, isospin diffusion in heavy-ion collisions \citep{tsang2009constraints}, nuclear masses \citep{kortelainen2010nuclear}, neutron skin thicknesses \citep{chen2010density} and electric dipole polarizability \citep{tamii2011complete, roca-maza2013electric}) constraining the symmetry energy. In particular, the recent PREX-II experiment \citep{Adhikari2021Accurate,reed2021implications} measured the neutron skin thickness of $^{208}\text{Pb}$ to be significantly higher than expected, implying a large value for $L$. 

Further studies of the symmetry energy are of great interest, but it is difficult for terrestrial nuclear experiments to access high isospin asymmetry similar to the state of matter in neutron stars. 
Neutron stars (NSs) are astrophysical compact objects and are the only places where matter reaches saturation density  and above on a macroscopic scale, and this -- combined with their extreme isospin asymmetry -- makes them ideal environments for studying the symmetry energy.
In particular, the composition of the elastic-solid NS crust is sensitive to the symmetry energy at around half saturation density \citep{steiner2008neutron,Oyamatsu2007Symmetry}, 
and the crust-core transition density 
is correlated with the slope and curvature of the symmetry energy below saturation density \citep{Ducoin2011Core,link1999pulsar}. Unlike terrestrial experiments, the challenge of using NSs to constrain the symmetry energy lies in identifying observational phenomena which allow us to probe the internal structure and composition of these compact stellar bodies.

Astrophysical observables that probe the overall size and compactness of the neutron star -- such as mass-radius constraints (see e.g. \citep{Raaijmakers:2021fk,Miller:2021qy}) or the tidal-deformability (\citep{Raithel:2018ai,Abbott:2018fe}) -- are mainly sensitive to the physics of the ultra dense inner core (see Fig. \ref{fig:violin} and Supplementary material). The core equation of state (EOS) is unknown, and may be controlled by the non-hadronic degrees of freedom \citep{weber2009neutron, miller2021astrophysical} that are unrelated to the nucleonic symmetry energy parameters.

Asteroseismology can probe the microphysical properties of particular regions of a neutron star, depending on where an asteroseismic mode is concentrated. Of particular interest for the symmetry energy is the quadrupolar crust-core interface mode ($i$-mode), which exists primarily at the crust-core transition and is restored by shear forces. As shear forces depend on the composition and depth of the crust, the properties of the $i$-mode contain information about the symmetry energy at $\sim$ half saturation density. Unfortunately, asteroseismic effects on gravitational-wave (GW) signals are typically weak, with the sensitivity of next-generation GW detectors required to observe the phase-shift due to mode resonances. However, multi-messenger astronomy may provide a way in which to measure the $i$-mode frequency: Resonant Shattering Flares (RSFs).

RFSs are short gamma-ray flares that occur when a normal mode of a magnetised NS is resonantly excited by the tidal field of the NS’s binary partner, such that the crust is strained beyond its elastic limit \citep{tsang2012resonant,tsang2013shattering,neill2022resonant}. In our previous work the $i$-mode was identified as a strong candidate for triggering RSFs, as it primarily oscillates at the crust-core transition
and is resonant $\sim$ seconds before merger. The natural frequency of a resonant mode can be precisely measured with coincident timing of a GW chirp and RSF, as the frequency of GWs from an inspiraling BHNS or NSNS binary when a RSF is triggered will be equal to the frequency of the resonant quadrupolar $i$-mode \citep{tsang2012resonant}. 
In \citet{neill2022resonant} we found that while not all NSs in binary mergers produce observable RSFs, there may still be several events per year where both GWs and a RSF are strong enough to be seen with current detectors, allowing this $i$-mode frequency measurement to be made.

In this Letter, we will investigate how much we can expect to learn about the symmetry energy parameters in the event that the $i$-mode frequency is measured using coincident RSF and GW timing.
First, we will outline how a Skyrme mean-field model parameterised by the nuclear symmetry energy parameters can be used to construct the NS equation of state (EOS) and composition. We will then perform Bayesian inference of this model's parameters using data from an injected multi-messenger RSF and GW detection. The results of this inference will be discussed in the context of constraints from terrestrial nuclear physics experiments.

\section{Nuclear model} \label{sec:model}

To calculate the EOS and composition of the crust we use an extended Skyrme energy density functional (EDF) to model the EOS in up to $1.5n_{\rm s}$. Full details on how we construct the EOS can be found in the suppplementary material and \citet{Stone:2007uq,Zhang:2016ww,newton2021nuclear}. We sample the parameter space of the symmetry energy by tuning the Skyrme parameters to give the required parameters $J$,$L$ and $K_{\rm sym}$ \citep{Newton:2013aa,balliet2021prior}. The remaining nuclear matter parameters are fixed at values of the Skyrme parameterization Sk$\chi$450 \cite{Lim:2017aa}. These EDFs are then used in a compressible liquid drop model (CLDM) to obtain the EOS, composition of the crust and the shear modulus required for the $i$-mode. \citep{balliet2021prior}. When the energy density favors uniform matter we transition to uniform $npe\mu$ matter using the same Skyrme. This crust-core transition density is known to be sensitive to the symmetry energy \citep{Oyamatsu2007Symmetry,Horowitz:2001aa,Xu2009nuclear,Xu2009locating}.

Moving deeper inside the star’s core the density increases to several times saturation density. In the core relativistic effects, and the likely transition to quark degrees of freedom in the inner core, mean the symmetry energy -- which assumes only nucleonic degrees of freedom -- becomes inapplicable \citep{Baym2018}. We then use a piecewise polytrope method \citep{Read2009Constraints,Read2009Measuring,Steiner2010equation,Steiner2013Neutron,Ozel2009Reconstructing,Ozel2010Astrophysical,Ozel2016Dense}
in the inner core; we attach two polytropes at 1.5$n_{\rm s}$ and 2.7$n_{\rm s}$ with polytropic indices ${n_1}$ and ${n_2}$ \citep[for details, see][]{Newton2018testing,neill2021resonant}.  
Low $n_i$ result in a stiff EOS, while higher $n_i$ give a softer EOS. Our NS EOS is thus characterized by 5 parameters: $J$, $L$, $K_{\rm sym}$, $n_1$, $n_2$.

\section{Bayesian inference of the model's parameters}\label{sec:inference}

To extract nuclear symmetry energy parameters from an $i$-mode frequency measurement, we use a Markov chain Monte Carlo (MCMC) method to perform Bayesian inference of our model's input parameters. In this section we outline our priors, likelihood functions and injected data.

{\bf Priors:}
We begin with uniform distributions over conservative ranges of the symmetry energy parameters consistent with that inferred from a variety of experimental nuclear data \citep{lattimer2014constraints}: $25 < J < 45 \text{ MeV}$, $0 < L < 200 \text{ MeV}$, $-600 < K_{\rm sym} < 200 \text{ MeV}$, while for both polytrope parameters we impose the bounds $n_i>0.001$ ($n_i<200$) to avoid having them go to zero (infinity) in the limit of maximally stiff (soft) EOSs.
Some regions of the resulting parameter space do not give viable NS models; where this occurs we simply set the prior probability to zero. Viability requires a crust and core stable with respect to small density perturbations and a causal EOS. Since $n_i$ has reasonable values over several orders of magnitude, and the region with $n_1>1$ and $n_2>1$ usually fails to produce a viable NS model, we use $\log_{10}(n_i)$ as parameters in our inference, not $n_i$. 

\begin{figure}
\centering
\includegraphics[width=0.5\textwidth,angle=0]{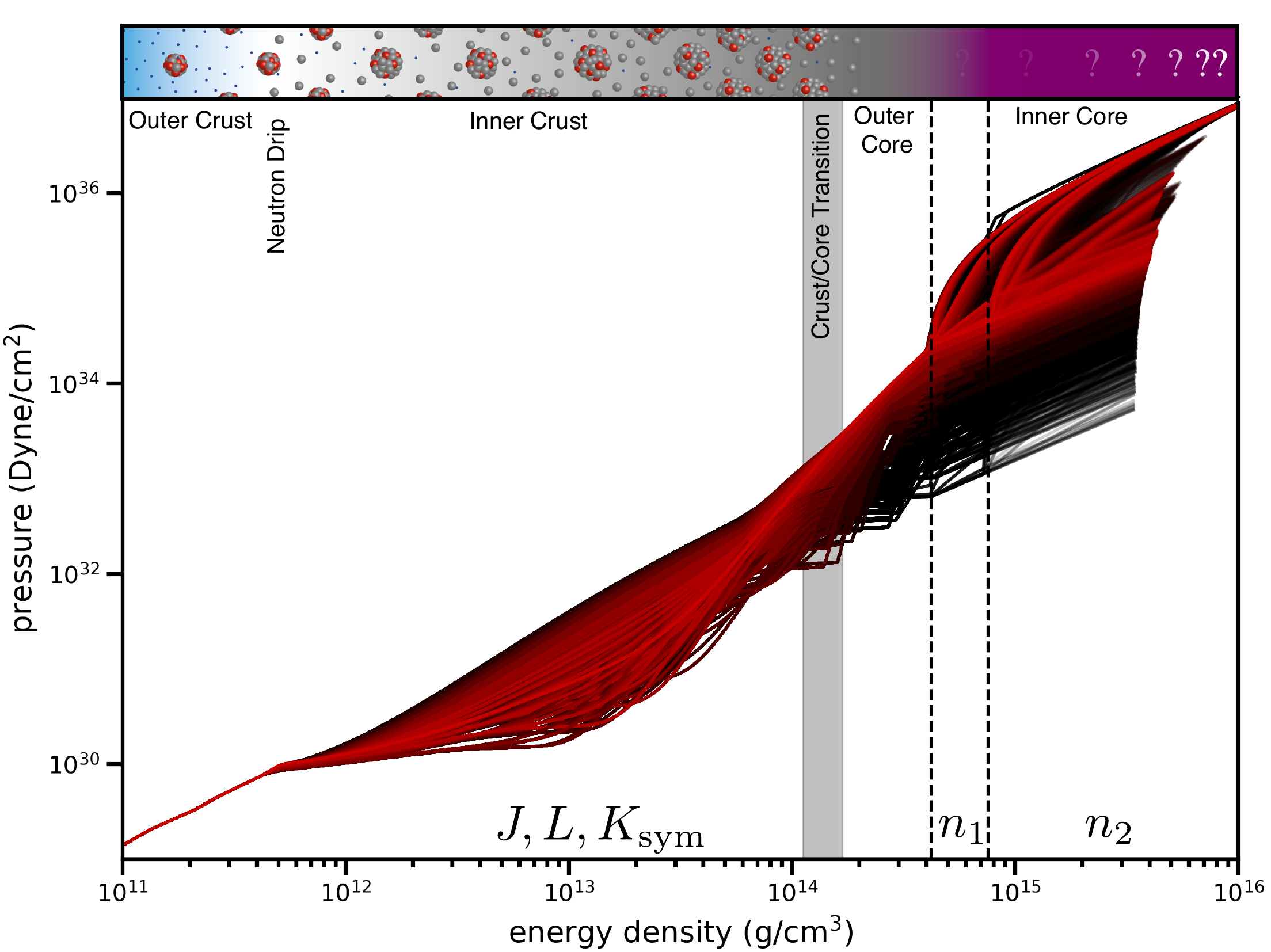}
\caption{NS EoSs produced over the full viable region of our parameter space, showing the range of $p(\rho)$ space covered. Dashed lines separate sections of the star that are affected by different parameters. 
 The outer crust EOS is taken from \citet{baym1971ground}, while the inner crust and outer core use an extended Skyrme EDF parameterized by the (isovector) symmetry energy parameters $J$, $L$, and $K_{\rm sym}$. Above $1.5 n_{\rm s}$ a simple piecewise polytrope is used to represent the unknown core physics, with piecewise transition at $2.7 n_{\rm s}$.  
We use the colours of the EOSs to show their $L$ values, with brighter red indicating higher $L$ in the range $0 < L < 200 \text{ MeV}$.}
\label{fig:EoSs}
\end{figure}

This uniform distribution over the viable parameter ranges is a relatively uninformative prior, allowing us to see the effects of various data on the inferred parameter values.
Figure~\ref{fig:EoSs} shows the range of EoSs that will be considered in our inference, where we have used the colours of the EoSs to indicate their $L$ values in order to show the importance of this parameter around the crust-core boundary.

{\bf Likelihood function and data:}

The frequency of the $i$-mode for a given NS EOS and composition can be calculated \citep[see][for details and approximations]{neill2021resonant} using the linearised relativistic pulsation equations of \citet{yoshida2002nonradial}.
By comparing this calculated $i$-mode frequency to the GW frequency observed during a RSF we can obtain the likelihood of a NS EOS and composition, and thus the model parameters used to generate them.

As there has yet to be a multi-messenger RSF detection, we inject one at $f_{\rm RSF}=250\text{ Hz}$ produced by a NS with mass $1.4 \text{ M}_{\odot}$.
There is no significance to this choice of frequency, except that it is mid-range for our models.
Over the $\sim 0.1\text{ s}$ mode resonance window \citep{neill2022resonant}, the GW frequency will sweep through a range of approximately \citep{tsang2012resonant}
\begin{equation}
\delta f\sim t_{\rm res}\frac{\partial f_{\rm GW}}{\partial t}\bigg\rvert_{f_{\rm RSF}}\sim3.7\text{Hz}\left(\frac{\mathcal{M}}{1.2\text{M}_{\odot}}\right)^{\frac{5}{6}}\left(\frac{f_{\rm RSF}}{100 \text{ Hz}}\right)^{\frac{11}{6}}, \nonumber
\label{eq:freq_spread}    
\end{equation}
\noindent where $\mathcal{M}$ is the chirp mass of the binary system \citep{cutler1994gravitational}. 
In a real multi-messenger detection it will not be clear when during the flare the exact resonance occurred, introducing an uncertainty of $\sim \delta f$ in the measured $i$-mode frequency. We therefore choose to have our injected data be a normal distribution around $f_{\rm RSF}$, with a conservative standard deviation $\delta f$. 

As a comparison to other astrophysical observables, we shall also consider the most significant NS measurements: mass-radius constraints from pulse-profile modelling of PSR~J0030+0451 and PSR~J0740+6620 \citep{riley2019NICER,riley2021NICER}, and the tidal deformability constraint from the GW170817 \citep{abbott2017gw170817,abbott2019properties}.
When including these data, the posteriors will be labelled with ``+Astro''.

\begin{figure*}
\centering
\includegraphics[width=0.95\textwidth,angle=0]{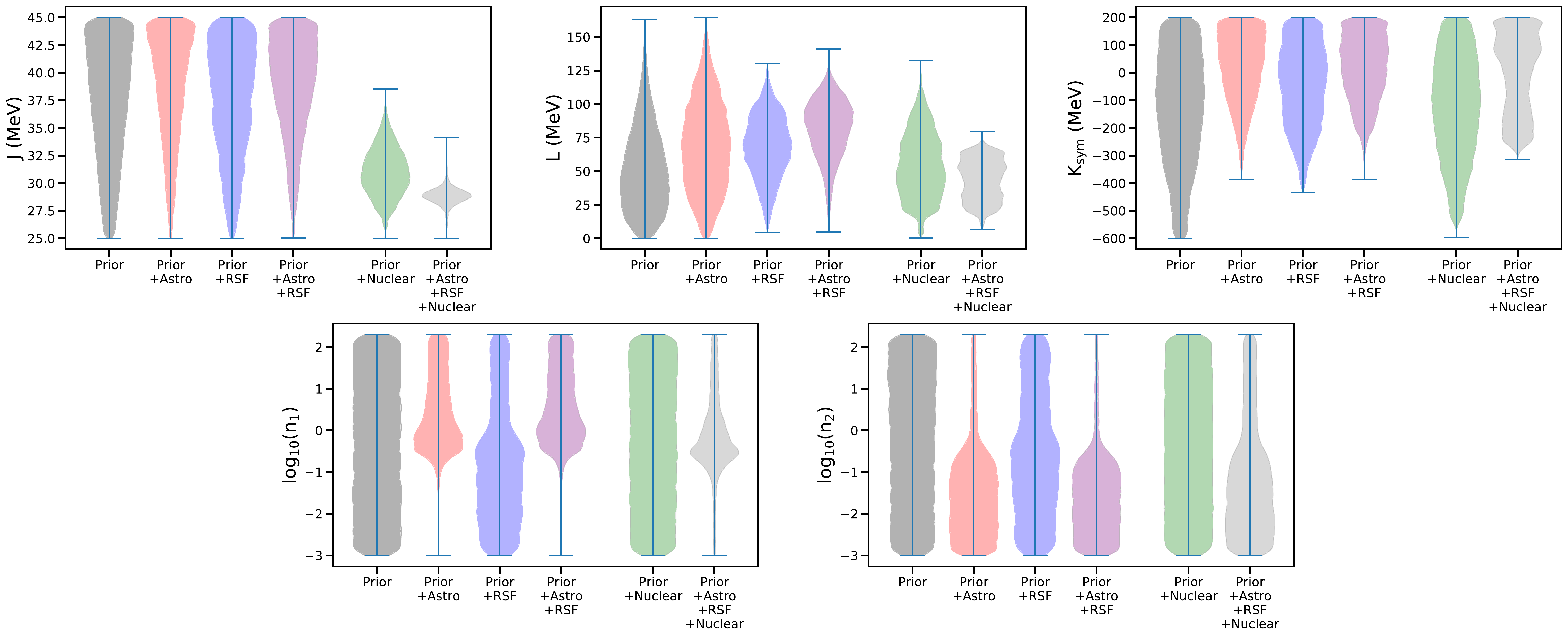}
\caption{Violin plot for various combinations of astrophysical and nuclear data. They all use the same uniform prior over the ranges of the parameters that produce viable NS EOSs (Prior), and the data included are: GW170817's tidal deformability and NICER's mass-radius measurements of PSR~J0030+0451 and PSR~J0740+6620 (+Astro), the i-mode frequency measurement from an injected RSF detection (+RSF), and the binding energies of various doubly magic nuclei (+Nuclear). This figure shows that multi-messenger timing of RSFs is particularly informative for $L$, and relatively insensitive to the NS core, while the other astrophysical constraints are mainly sensitive to the core parameters.}
\label{fig:violin}
\end{figure*}

{\bf Posteriors:}
We use our priors and likelihood functions with the \textit{emcee} MCMC python module \citep{foreman-mackey2013emcee} to perform Bayesian inference of our EOS's parameters. 
Figure~\ref{fig:violin} contains violin plots for each of our five EOS parameters, showing the 1D probability distributions for our: uniform prior for viable EOSs, posterior for the astrophysical data, posterior for the injected RSF, and posterior using both the astrophysical data and injected RSF. We also show the posterior using data from nuclear binding energies alone, and our full posteriors when this nuclear data is included (for full corner plots, see Supplementary Material). 
The main result of our inference can be seen by comparing the 2nd and 4th bands: the $i$-mode frequency measurement from multi-messenger GW and RSF detection significantly improves the constraint on $L$ compared to only using core-dependant astrophysical observables, while being relatively insensitive to the parameters describing the NS core ($n_1$, $n_2$ and $K_{\rm sym}$). The astrophysical data has a large effect on the inferred values of these core parameters but has a much smaller effect on $J$ and $L$. Finally, nuclear masses strongly constrain $J$ and have a smaller effect on $L$.

\section{Discussion}\label{sec:discuss}

\begin{figure*}
\centering
\includegraphics[width=0.95\textwidth,angle=0]{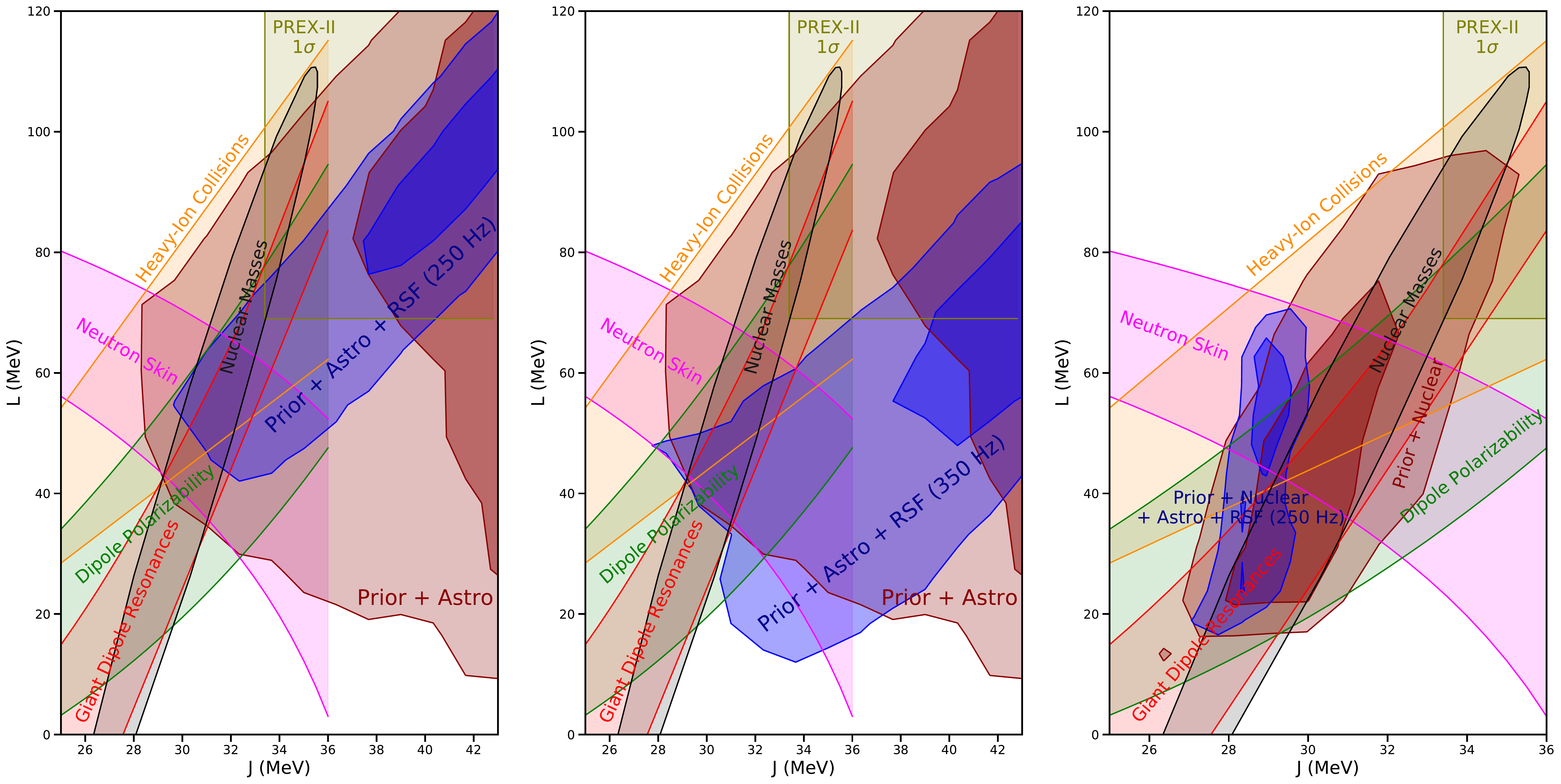}
\caption{Various experimental and observational constraints on the nuclear symmetry energy parameters \citep{trippa2008giant,tsang2009constraints,kortelainen2010nuclear,chen2010density,tamii2011complete,roca-maza2013electric}, similar to figures in \citet{lattimer2014constraints,reed2021implications}. The three panels show different RSF posteriors. Left: the red regions are the $1$ and $2\sigma$ posteriors using just the core-dependant astrophysical data, which become the blue regions when a RSF injected at $f_{\rm RSF}=250\text{ Hz}$ is added to the data. Middle: the same, but with the RSF at $350\text{ Hz}$ instead. Right: The red regions are the posteriors for just nuclear data, and the blue regions are for nuclear, astrophysical and $250\text{ Hz}$ RSF data. RSFs provide strong constraints on $L$, with the specific $L$ values determined by the GW frequency at which the flare is detected. Nuclear data strongly constrains $J$, making it complementary to RSF data.}
\label{fig:constraints}
\end{figure*}

In Figure~\ref{fig:constraints} we show the the correlation between $J$ and $L$ in the RSF posteriors. Our $1$ and $2\sigma$ posterior regions in $J$ and $L$ are plotted alongside the results of various nuclear experiments, including the recent PREX-II result. Several of these experimental constraints come via measurements of the neutron skin thickness -- the differences in the average radii of protons and neutrons in nuclei -- which are closely related to $L$ in particular. We see that a single coincident RSF and GW detection may provide a comparable constraint (at $2\sigma$) to heavy ion collision \citep{tsang2009constraints} and dipole polarizability \citep{roca-maza2013electric} experiments, and that while $J$ alone is not strongly constrained, its value is highly dependent on $L$. Since the injected $i$-mode frequency, $f_{\rm RSF} = 250$ Hz, was arbitrary, we also show the posterior for an injected frequency of $f_{\rm RSF} = 350$ Hz, which results in a shift of the $J$-$L$ constraint region of $\sim -20$ MeV in $L$. 

These posteriors include no information from nuclear experiment. The requirement that EOS models be  physically-viable results in high $J$ being favored by our priors, as higher values of $J$ allow wider ranges of the other parameters to be viable. Tidal deformability and mass-radius relationships do not contain much information about less dense regions of their stars, 
and so the bias towards high $J$ is preserved when these astrophysical data are included. However, these priors are sufficient to show that RSFs alone can give us strong constraints on $L$ and the $J$-$L$ relationship and provide little new information about the NS core.

Including well-constrained nuclear masses in our data significantly constrains $J$ as shown on the right of Figure~\ref{fig:constraints}.
The $i$-mode frequency is mainly dependent on the properties of matter around $0.5n_{\rm s}$, and so its posteriors show significant degeneracy between $J$ and $L$. Nuclear masses mainly probe the bulk EOS around saturation density, so are less dependent on the slope and curvature (as can be seen from the `Prior+Nuclear' bands on figure~\ref{fig:violin}). Including nuclear masses therefore breaks the $J$-$L$ degeneracy in our RSF posteriors, giving us the significantly smaller $L$ range shown in the final band of figure~\ref{fig:violin}'s $L$ plot. The complementary nature of these data illustrates the importance of probing matter at several different densities, and our results show that the NS crust is an important source of sub-saturation constraints. 

\section{Conclusions}

The nuclear symmetry energy is an important property of neutron-rich nucleonic matter which plays a crucial role in determining the properties of neutron stars. However, it is uncertain if such nucleonic equations of state are valid at all densities within a neutron star, or if non-hadronic degrees of freedom become important beyond a few times nuclear saturation density \citep{Baym2018}. Thus, astrophysical observables which depend mainly on core properties -- such as masses and radii, tidal deformability, or f-mode frequency -- may not be reliable probes of nuclear symmetry energy. Instead, we argue that astrophysical symmetry energy constraints should focus on observables that probe regions where we are confident that nucleonic physics is dominant. One such observable is the quadrupolar crust-core interface mode which can be excited by tidal resonance, and depends primarily on the physics of the crust-core transition region. Unlike other resonant asteroseismic modes, the frequency of of the $i$-mode can be precisely measured by coincident timing between a gravitational-wave chirp, and a gamma-ray resonant shattering flare \citep{tsang2012resonant,tsang2013shattering,neill2021resonant,neill2022resonant}. While not all gravitational-wave chirps will be accompanied by such a flare, multi-messenger events may be common enough for these measurements to be made somewhat frequently with current detectors \citep{neill2022resonant}.

In this work we have used an equation of state which couples an extended Skyrme EDF -- parameterised by the symmetry energy parameters $J$, $L$ and $K_{\rm sym}$ -- with a simple piecewise polytrope at higher densities which represents our uncertainty of the nature of matter in the core. Using this parameterized EOS along with a compressible liquid drop model for crust composition, we have explored the power of RSFs as a tool to constrain the nuclear symmetry energy. By injecting representative RSF detections we used a Bayesian analysis to determine the posteriors for the parameters in this equation of state. Using conservative priors and pessimistic data, a single multi-messenger RSF and GW detection may be an equally strong tool for constraining the first two nuclear symmetry parameters as some terrestrial collider experiments (see Figure~\ref{fig:constraints}, left and middle). Additionally, including nuclear mass data in the inference further reduces the posterior region of an RSF detection 
(Figure~\ref{fig:constraints}, right). 

A clear picture emerges: different observables are sensitive to different densities: nuclear masses constrain the crust through $J$, resonant shattering flares constrain the crust and outer core through $L$, and radius and tidal deformability data constrain the outer and inner core through $K_{\rm sym}$, $n_1$ and $n_2$.

Our results do include model dependencies: although we allow a wide exploration of the parameter space of the nuclear EOS, the choice of Skyrme model may still restrict the density dependence, although to much smaller extents when one goes beyond the second-order $K_{\rm sym}$ term. The CLDM contains surface parameters that were fit to calculations of crust nuclei from a relatively small number of EDFs, introducing a possible model dependence that will be explored in future works. None of these are expected to qualitatively change the message of this paper.

Constraints from RSFs on NS structure will be complementary to those from tidal deformability or mass-radius constraints, allowing for relatively independent probes of crust and core. With the upcoming fourth LIGO/Virgo observing run, the first multi-messenger RSF detection may occur in the near future. Looking ahead to next-generation GW interferometers, the number of such detections will likely increase significantly; we shall investigate the advantages of having multiple detections in future work.

\begin{acknowledgments}
\emph{Acknowledgements}
DN is supported by a University Research Studentship Allowance
from the University of Bath. WGN and RP were supported by the NASA grant 80NSSC18K1019.
\end{acknowledgments}

\software{emcee \citep{foreman-mackey2013emcee}, skyrme\_rpa \citep{Colo:2013bh}}

\bibliography{RSFSymmetry}{}
\bibliographystyle{apsrev4-1}

\end{document}